# Choosing Collaboration Partners.
## *How Scientific Success in Physics Depends on Network Positions*

Raphael H. Heiberger*[1] & Oliver Wieczorek[+]

*University of Bremen*

[+]*Otto-Friedrich University Bamberg*

Physics is one of the most successful endeavors in science. Being a prototypic "big science" it also reflects the growing tendency for scientific collaborations. Utilizing 250,000 papers from ArXiv.org – a pre-publishing platform prevalent in Physics – we construct large co-authorship networks to investigate how individual network positions influence scientific success. In this context, success is seen as getting a paper published in high impact journals. To control the nested levels of authors and papers and to consider the time elapsing between working paper and prominent journal publication, we employ multi-level event-history models with various network measures as covariates. Our results show that the maintenance of a moderate number of persistent ties is already crucial for scientific success. Also, even with low volumes of social capital Physicists who occupy brokerage positions enhance their chances of publishing articles in high impact journals significantly. Surprisingly, inter(sub)disciplinary collaborations decrease the probability of getting a paper published in specialized journals for almost all positions.

Acknowledgements:
We would like to gratefully acknowledge comments and discussions with Christian Baier, Richard Heidler, Pia König, Richard Münch, Brigitte Münzel and Daniel Wieczorek.

[1] Corresponding author: Raphael H. Heiberger, University of Bremen, Institute for Sociology, Mary-Somervile Str. 9, 28205 Bremen, Germany, +49 421 218 58528, rhh@uni-bremen.de

# 1. Introduction

Science is a collective human enterprise designed to produce new knowledge and provide solutions for societal, medical, technological and environmental challenges. Collaborations with colleagues are among the basic requirements of doing science (e.g. Brint, 2001; Barabási et al., 2002; Leahy and Cain, 2013). Co-authorships improve the chances to produce papers (Lee and Bozeman, 2005) and increase the flow of knowledge between disciplines (Harris, 2010; Uzzi et al., 2013; Rawlings et al., 2015) as well as between academia and industry (Boardman, 2009; Ponomariov and Boardman, 2010; Witherspoon, 2012). In a changing academic environment characterized by global competition (Marginson, 2006; Adams, 2012; Münch and Schäfer, 2014), financial strain (Smith-Doerr, 2005; Berman, 2008; Leslie et al., 2012) and the need to raise one's own performance by publishing in high-impact journals (Bornmann and Daniel, 2006: 45f.; Münch, 2014: pp. 222–233) the number of co-authorships has risen steeply in the last decades (Glänzel, 2001; Moody, 2004; Newman, 2004; Rushforth and De Rijke, 2015).

Recognition of and working with peers is therefore one of the major challenges and central requirements for scientists. But what are the *characteristics within collaboration networks that influence scientists' outreach*, i.e. which positions in networks raise the chance of getting a paper published in journals possessing a high impact in their specific field of work? Although many researchers have been engaged with scientific relationships and the resulting abundance of studies about scientific networks, we see three important gaps in the existing literature in regard to this question: (a) Unlike other studies, we link structural properties of collaboration networks empirically to being successful or not. Especially failure in the process of publishing has not yet been operationalized as a meaningful counterpart of being successful. (b) Longitudinal networks have been the subject of a growing number of studies in recent years (Snijders et al., 2010). However, the effect of persistent relationships is relatively unclear since there is "almost no research on the stability of interpersonal relationships" (Burt, 2002, p. 343). Burt´s argument still seems valid, although first evidence exists for the crucial role of "ties that last" (Dahlander and McFarland, 2013). (c) Interdisciplinarity is commonly interpreted as a pivot for both junior researchers (Jacobs and Frickel, 2009) and senior scientists (Leahy and Moody, 2014; Ylijoki, 2014). It is particularly emphasized by scholars conducting research on scientific innovation (Levitt and Thelwell, 2009; Uzzi et al., 2013; Bozeman et al., 2016). Yet, collaborations between *and* within (sub-)disciplines have so far not been addressed in regard to their influence of getting papers published.

Our paper concentrates on one of the arguably largest, best funded and most successful endeavors in the history of science: modern Physics (Kragh, 2002). It accounts for the development of many essential technologies of today´s world, for instance, the original electronic communication networks that led to the emergence of the internet (Campbell-Kelly, 1987; Leiner et al., 2009). Often regarded as "big science" (e.g. Galison, 1992), it is characterized by huge projects relying on complex networks in order to facilitate scientific breakthroughs (Wuchty al., 2007) and to handle the massive amount of data produced, e.g. by particle colliders or space-telescopes (Bodnarczuk and Hoddeson, 2008; Seidel, 2008).

The necessity to collaborate makes Physics an ideal subject to employ network analysis. This is even more the case since the field of Physics possesses a central publication platform – ArXiv.org – on which most papers are pre-published. Uploading a working paper, however, does not tell whether it is ultimately published in a well-recognized journal. Our analysis aims to explain how network positions influence the chances of achieving in terms of prominent publications. To this end, the paper is structured as follows: First, we discuss concepts of scientific success to clarify why publications in high impact journals constitute the "fundamental unit of success". Our main hypotheses are subsequently formulated as individual network properties that should theoretically influence the probability of getting a paper published. Hereafter, we introduce our ArXiv data set, provide some information about the platform itself and operationalize the derived hypotheses. In the main part of the paper, we analyze the network conditions for being successful in Physics by combining methods of quantitative network analysis with event-history analysis. This is only possible since the ArXiv data provide the time from uploading to publication in high impact journals (or not). Finally, we discuss our results in regard to existing studies as well as its potential limitations and formulate implications for future research.

## 2. What is Scientific Success?

Success in academia is either linked to individual researchers or disciplines on a larger scale. On a personal level, success is often seen as being appointed professor (Jungbauer-Gans and Gross, 2013), engaging in technology transfer with companies (Boardman and Ponomariov, 2007), or gathering funds (Feinberg and Price, 2004; Bromham, Dinnage and Hua 2016). Contributing to the scientific endeavor, scholars previously defined individual success as citations per paper, the number of published journal articles or a combination of both (Börner et al., 2005; Jansen et al., 2010; Acuna, Allesina and Kording, 2012; Abbasi et al., 2012). In contrast, an academic discipline is regarded as successful if it is able to answer fundamental questions, to support industrial collaboration and the transfer of ideas (Stuart and Ding, 2006; Welsh al., 2008; Boardman, 2009), to stimulate economic growth (Slaughter and Rhoades, 2005; Franzoni and Lissoni, 2009; Etzkowitz and Viale, 2010; Lam, 2010), to receive funding by institutionalized competition (Münch, 2013; 2014, pp. 126 – 165), or if it is declared relevant by politics (Slaughter and Cantwell, 2012).

In both perspectives, however, "success" is strongly linked to research output. Though there are many possible ways to be considered successful in the world of science, there is one basic requirement: publications (e.g. Crane, 1972; Merton, 1968; Münch, 2014). The implicit imperative behind that is *to publish or perish*, i.e. to produce new insights that others (mainly the respective peers) approve as such (e.g. Merton, 1968; Münch, 2014, p. 15). Certainly, there are large differences in epistemological cultures across countries and disciplines (Whitley, 2000). In most of the Humanities, for instance, the main communication still relies on monographs, while books in Physics appear only as textbooks for students to sum up existing disciplinary knowledge or as highly specialized lecture notes intended for a very limited circle of readers. Instead, new findings are spread via articles in specialized journals. Peer-reviewed publications in relatively far reaching prominent journals (i.e. "high impact journals") play a central role in the scientific reward system (Rushforth and De Rijke, 2015). Getting a paper published in those journals is increasingly relevant for researchers' academic careers, not least because publications become an increasingly crucial part of academic ranking systems (Willmott, 2011; Lee, Pham and Gu, 2013; Pusser and Marginson, 2013; Osterloh and Frey, 2015). Therefore, publications in high impact journals are considered a success in a dual sense: being recognized by colleagues while meeting the demands of performance measurements.

We utilize this pivot of scientific careers as "fundamental unit of success" and, hence, as our dependent variable, because (1) publishing is a necessary condition for being recognized by colleagues and gaining merits in the scientific community. Since articles distributed in high impact journals find a large, yet specific audience, the chances for a scholar's ideas to be cited are higher. The reputation gained from respective journals also enables the scholar to move up on the career ladder. (2) Focusing on the chances for publication in high impact journals allows us to define the possibility of "failure", that is not having a working paper published at all or in rather peripheral journals only. (3) Publications allow us to take into account the time passed from putting a paper online on ArXiv to its final acceptance by a leading journal. The time span from submission to publication is especially important in Natural Sciences like

Physics, since they are characterized by intense competition to solve theoretical riddles resulting in a high time pressure to be the first to publish findings. (4) Finally, using publications as central assessment for success reflects the scientific discourse in a discipline. In the long run, the ability to generate new knowledge and the acceptance of innovative ideas are decisive for the success of Physics as a discipline. Investigating the individual chances to get published as the conditions of success for the whole discipline is a frequently neglected indicator on the input side of the academic community, which helps to understand the underlying processes that precede this basic unit of scientific communication.

### 3. What Aspects of Networks Influence Scientific Success?

The widespread practice of writing papers together and the public availability of data make scientific relations, in our case within the discipline of Physics, a prominent field for network researchers (Freeman, 2004; Light and Moody, 2011; Newman, 2001). In general, network structures condition and frame individual opportunity spaces, identities and actions (e.g. White, 1992). Previously investigated scientific relationships comprise citations between articles (e.g. Gentil-Beccot et al., 2009), collaborative ties between institutions (e.g. Ponds et al., 2007) or countries (e.g. Heiberger and Riebling, 2015) as well as within national research systems (e.g. Kronegger et al., 2011). Arguably the most common scientific relationships are co-authorships (e.g. Melin and Persson, 1996). We will concentrate on those.

From a genuine network analytical view, scientific relationships in terms of co-authorships follow a logic known as *preferential attachment* (Barabási et al., 2002; Barabási and Albert, 1999). This means that the possibility of tie formation is constrained by the number of a node's previously realized ties. Since maintaining many ties consumes large amounts of time, scholars at the center of a field must be selective in their choice of collaborators. According to preferential attachment, they can afford to be picky because new, peripheral researchers in a network should connect mostly to them as researchers at the center.

This asymmetry is reinforced by mentor/mentee relations between elder and younger scholars and rivalry among different groups of scholars (Baier and Münch, 2012; Fochler et al., 2016), leading to segregation among groups that resemble "small worlds" (Watts and Strogatz, 1998; Amaral et al., 2000; Newman, 2001a; 2001b; Cohen and Havlin, 2002). This structure of whole networks is well known in Natural Sciences.[1] In this paper, however, we focus on less understood network effects that influence individual chances of getting a paper published in high impact journals and subsequent patterns between success and individual positions in the network. To this end, we derive network attributes of Physicists that we assume to condition the probability of getting a paper published in Physics' leading journals.

### 3.1. Node Positions

One possibility to understand the complex figurations of scientists linked by co-authoring a paper is to examine the individual position of each node in relation to the rest of the network (Wassermann and Faust, 1994). In social network analysis, having many ties (e.g. friendships, business partners, etc.) is commonly seen as "better" (Freeman, 1979). The question as to what "better" means in each context is rather difficult to answer and, for instance, discussed in greater detail by Borgatti (2005). Within the framework of scientific collaborations in general and in Physics in particular, the interpretation is that having many research partners is evidence of a scientist's general attractiveness and provides greater opportunities for further collaborative work and recognition by peers. Thus we formulate our first hypothesis:

**H1a**. The more collaborations a Physicist maintains, the more likely is his/her scientific success.

However, not all ties are equal. The impact of network relations is often time-lagged and sustaining existing collaborations requires other strategies than forging new ones (Dahlander and McFarland, 2013). It makes a huge difference whether collaboration takes place between people who have known each other already or whether it is their first work together. As Brint (2001, pp. 399–400) suggests, it is crucial to have "a small circle of people whom [the researchers] talk to regularly about their work". In addition, the findings of Leahy and Cain (2013, pp. 936–941) point out that being part of a stable network is highly important for being successful in academia. This is due to the fact that important social mechanisms – like trust – need time to develop (Krackhardt, 1992). For instance, honest and direct discussions spanning hierarchies and intellectual boundaries are more likely if collaboration partners know each other well through repeated co-authorships. To consider the persistence of scientific

---

[1] There is much less research concerning the structure of Social Sciences. For instance, collaboration patterns in sociology resemble a structural cohesion network with a large, strongly connected network component (Moody, 2004; Moody and White, 2003; White and Harary, 2001).

collaborations as a condition of scientific success we formulate a second hypothesis that also grounds on the level of a node's degree but considers the duration of ties:

**H1b**. The more persistent ties a Physicist has, the more likely is his/her scientific success.

One of the most important individual resources incorporated in human relationships is social capital (e.g. Coleman, 1990; Lin, 2002). As Bourdieu (1986) famously pointed out, social capital does not only consist of the number of connections of an ego but also with whom he or she maintains these connections. In network terms this means that more ties with partners who also maintain many relations in their turn yields higher social capital (Burris, 2004). The acquisition of interpersonal resources is necessary to be accepted in any privileged status group. At the same time, membership in a group increases along with the group's exclusiveness and the resources it controls (Bourdieu, 1984). In scientific relationships social capital rests on the "prestige principle" (Han, 2003) and plays a major role in individual scientists' collaborations (Bozeman and Corley, 2004). Hence, co-authorships with prominent researchers provide access to higher hierarchic levels, increase one's own academic prestige and should make it easier to publish articles in high impact journals:

**H1c**. The higher a Physicist's social capital, the more likely is his/her scientific success.

It has been observed that the distribution of social capital in scientific collaboration networks is highly skewed (Merton, 1968; Burris, 2004; Li et al., 2013). The resulting networks possess a tightly connected center with many ties and a periphery characterized by few ties. Following Elias and Scottson (1994), we can relate this typical structure to the distinction between groups of established researchers and peripheral outsiders. Distinction between established and peripheral Physicists might also be at work in the case of preferential attachment since scientific prestige is associated with more central positions (e.g. Newman, 2001). Hence, preferential attachment is a mechanism that stabilizes centers within networks over time and lowers the chances of outsiders to accumulate social capital. At the same time, such clusters increase the flow of redundant information (Moody and White, 2003; Burt, 2004).

In this manner, the notion of social capital is often seen to emphasize stability and reproduction of network formations by maintaining the inherited positions in a network (Walker et al., 1997). Ronald Burt (1995), however, developed a complementary view on social capital. He emphasizes the opportunities created by brokerage positions that connect otherwise unconnected parts of a network. Brokers foster the flow of ideas and connect heterogeneous actors across *structural holes* to reveal otherwise hidden options (Burt, 2004). The opposite of such brokerage positions are formations of nodes whose contacts are highly connected to each other and, hence, contain mostly redundant information in highly constrained networks. Burt´s argument has originally been developed in economic settings in order to explain entrepreneurial innovation but has been observed in various social contexts, e.g. only recently in the creation of video games (De Vaan et al., 2015). For scientific relationships it has proven to be one guiding principle in the assemblage of teams by including a certain fraction of newcomers with non-redundant information (Guimerà et al., 2005). The enhancement of creativity by bridging structural holes leads to our last hypothesis in regard to node positions:

**H1d**. The less constrained a Physicist's relationships are, the more likely is his/her scientific success.

### 3.2. Interdisciplinarity

Apart from an individual scientist's network position we assess the interdisciplinarity of research collaborations as being an important influence factor for success. This argument is tightly connected to the aspect of bridging structural holes since spanning bridges across disciplinary boundaries should lead to the import of new ideas. Increasingly, research across disciplinary boundaries is a requirement for high impact research and is seen as an important source for fruitful ideas in science and other creative enterprises (Rawlings et al., 2015; Uzzi et al., 2013). Current "big science" undertakings in Physics (e.g. at CERN) illustrate the need and necessity for collaboration in order to work towards the solution of fundamental problems such as gathering evidence for the existence of the "Higgs particle". Thus, in other words, the basic idea behind interdisciplinary collaboration efforts is that such teams effectively combine knowledge from different areas:

**H2a**. If a Physicist's collaboration project is spanning multiple scientific fields, it enhances his/her chances of scientific success.

Since Physics is arguably one of the largest and most differentiated scientific fields, we extend interdisciplinarity to collaborations between different subfields in Physics. The argument holds that such efforts span intellectual boundaries and respective epistemological cultures at the intersection of subdisciplines (Heidler, 2011; Leahy and Cain, 2013). Under such collaborations we subsume, for instance, collaborations between Nuclear and Particle Physicists. Thus we formulate following hypothesis:

**H2b**. If a Physicist's collaboration project is spanning multiple fields within Physics, it enhances his/her chances of scientific success.

### 3.3. Interaction Effects

*Constraint x Degree*

As mentioned in *H1d*, brokerage positions are not necessarily occupied by nodes featuring the greatest number of relationships in a network but rather those spanning structural holes. While high levels of degrees point to central nodes, the non-redundancy of these relations emphasizes the bridging of intellectual boundaries. However, an interplay between both is reasonable since many collaborations and, hence, a certain attraction for scientific partners should also increase the probability to connect nodes with different backgrounds. To consider Physicists' number and constraint of relations *together* we establish the following hypothesis:

**H3a.** The less constrained the relationships of Physicists are and the more relations a Physicist has, the more likely is his/her scientific success.

*Constraint x Social Capital*

Structural holes as understood by Burt (1995, 2004) stress a different aspect incorporated in relationships as the "classical" approach to social capital represented by Coleman (1990) and others. High amounts of social capital (i.e. many partners with many relations) can be seen as an indicator of scientific reputation (Burris, 2004). The interaction with social capital provides us with information about the strength of distinction (Bourdieu 1984) of the cluster the respective Physicist takes part in as well as the social closure of more or less centered Physicists in the collaboration network in line with the established/outsider hypothesis of Elias and Scottson (1994). Both are associated with social capital on the individual level as defined by Bourdieu (1986) and Lin (2002). Therefore, the interaction between constraint and social capital allows us to investigate how strongly social capital is associated with brokerage as discussed by Burt (2004). Especially in Physics, high reputation normally goes hand in hand with leading large research teams. If a "big shot" uses his or her reputation to collaborate with Physicists from outside the own working group, the flow of ideas should be non-redundant and have low constraints:

**H3b.** The less constrained the relationships of a Physicist are and the higher the social capital is, the more likely is his/her scientific success.

*Constraint x Interdisciplinarity*

Scientific disciplines structure academia and reflect intellectual proximity (Jacobs and Frickel, 2009). In the context of collaboration networks, structural holes should be especially present when bridging those boundaries. In order to avoid redundant information (Burt, 2004) and to mobilize social capital located in other subfields of science, we expect Physicists to actively seek collaborations in other subdisciplines or scientific disciplines. These interactions rely on two distinctions: First, the distinction between scientific disciplines and their respective discourses and cultures (Whitley, 2000) and second, the distinction between those established in the center and outsiders in the periphery (Elias and Scottson, 1994) We can therefore assume that when working with colleagues from other specializations Physicists who are not specialized in those areas utilize their knowledge and thus receive new information. This basic principle of division of labor should be particularly powerful if scientists occupy bridging positions *between* Physics and other disciplines, and also *within* different branches of Physics. Using interaction effects we can examine both cases and posit:

**H3c**. A Physicist who is less constrained and works on collaboration projects combining Physics with other disciplines is more likely to achieve in scientific terms.

**H3d**. A Physicist who is less constrained and works on collaboration projects spanning multiple fields within Physics is more likely to achieve in scientific terms.

## 4. The Structure of ArXiv

ArXiv[2] was established in 1991 and was originally an automated email server. It was constructed to serve about 100 submissions per year, assumed to stem from a then small subfield of High-Energy Physics. The system's benefit as an actual archive and, even more, as a way of global distribution of research ideas was quickly appreciated by the editorial directors of both major physical associations in the US, the American Physical Society and the Institute of Physics Publishing. The rapid progress of the internet during the 1990s led to a sheer explosion of the pre-refereed distribution of scientific papers in physics and related disciplines like Mathematics or Computer Science (Ginsparg, 2011). Currently, the total number of downloaded papers exceeds 700 million, with Physicists from all over the world and from all subdisciplines uploading their working papers on ArXiv (ArXiv, 2016a).

---

[2] The capital X in ArXiv is derived from χ of the TeX language and its first URL xxx.lanl.gov (Ginsparg, 2011).

Though ArXiv is by design an open-access platform for the free and rapid distribution of knowledge, submissions have to undergo some moderation managed by outstanding members of the community who are assembled in the ArXiv scientific advisory board (ArXiv, 2016b). The library rejects, for instance, inappropriate topics or formats as well as duplicates or plagiarized material.

The success of the ArXiv system has been recognized in a number of meta-analyses. Henneken and colleagues find that the most important papers in Physics and Astronomy appear first on ArXiv (Henneken et al., 2006). In High-Energy Physics, ArXiv preprints are also more frequently read than other articles (Gentil-Beccot et al., 2009). In some fields of Physics, virtually all articles are published on ArXiv first, especially in Astrophysics, Nuclear and Particle Physics as well as in High-Energy Physics. In Mathematics or Computer Sciences these numbers are much lower (Lariviere et al., 2013). In total, we assume that ArXiv represents major parts of the field of Physics since it is the de facto standard in most subareas.

## 5. Data and Methods

### 5.1. Collaboration Networks of Physicists on ArXiv

We retrieved the ArXiv data from the official API using Python and downloaded all articles related to Physics. The Application Programming Interface contains several instances of meta-information from each paper, and most importantly, the authors. The co-authorship networks are derived as follows: Two scientists, *i* and *j*, form a tie if they have co-authored a paper within a given year *t*. Let the number of edges between *i* and *j* be $A_{i,j}^t$, which will be 0 if no link exists, 1 if they have written one paper together and more than one if there are multiple edges. For *n* vertices this can be represented in an $n \times n$ adjacency matrix $\boldsymbol{A}^t$. The underlying assumption is that most people who have written a paper together will know each other quite well (e.g. Newman, 2001).

However, in Physics papers may have several hundred authors. For this reason, we weighed the ties $A_{i,j}^t$ according to the total number of authors per article. Thus a tie between two authors who wrote a paper together is two times stronger than a paper authored by four scientists. Assuming that *i* and *j* have participated in both articles of that sort, then $A_{i,j}^t$ would be ¾ ($\frac{1}{2} + \frac{1}{4}$). The intention behind the weighting scheme is that fewer authors mean stronger contacts between participants.

Another potential bias in constructing the networks are spelling mistakes. Even though the names are entered by the authors themselves we discover various spellings of the same name in the raw data insofar as first names are sometimes abbreviated and sometimes are not. To avoid duplicates we have reduced all first names to a single letter. Since we are interested in individual network positions of Physicists, we have also excluded all corporations, groups and teams as authors. Authors collaborating only with such groups have also been deleted. This procedure allows us to focus on direct collaboration in line with Leahy and Cain (2013) and avoid the fallacy of authors with an extremely high amount of publications each year (i.e. exceeding 100 publications per year), which seems possible only by cooperating with such large teams.

In total, our original data set comprises of 301,989 papers spanning from 2001 to 2014. Albeit our data starts in 2001, we decided to exclude data from the year 2001 in order to calculate the effect of persistent ties as defined in the following section. Also, the final sample concentrates on the years until 2011 to allow papers three years to get published. Eventually, our data includes 245,432 papers.

Figures 1 and 2 give an impression of the structure of the field of Physics in 2002 and 2011. By simply comparing both figures we can identify the growth of the field, which is in accordance with other observers (e.g. Leydesdorff and Wagner, 2008). At first, the field consists of loosely connected researchers and only few tightly connected groups which are illustrated by colors derived from the "Louvain" implementation of the modularity algorithm (Blondel et al., 2008; Clauset et al., 2004). In 2011, the structure has clearly changed. We see a far more differentiated field, with tightly connected clusters dominating the picture. The relationships of those groups extend like kraken arms into the main field. Thus the production of papers has increased massively over the last decade resulting in a more modularized network topology.

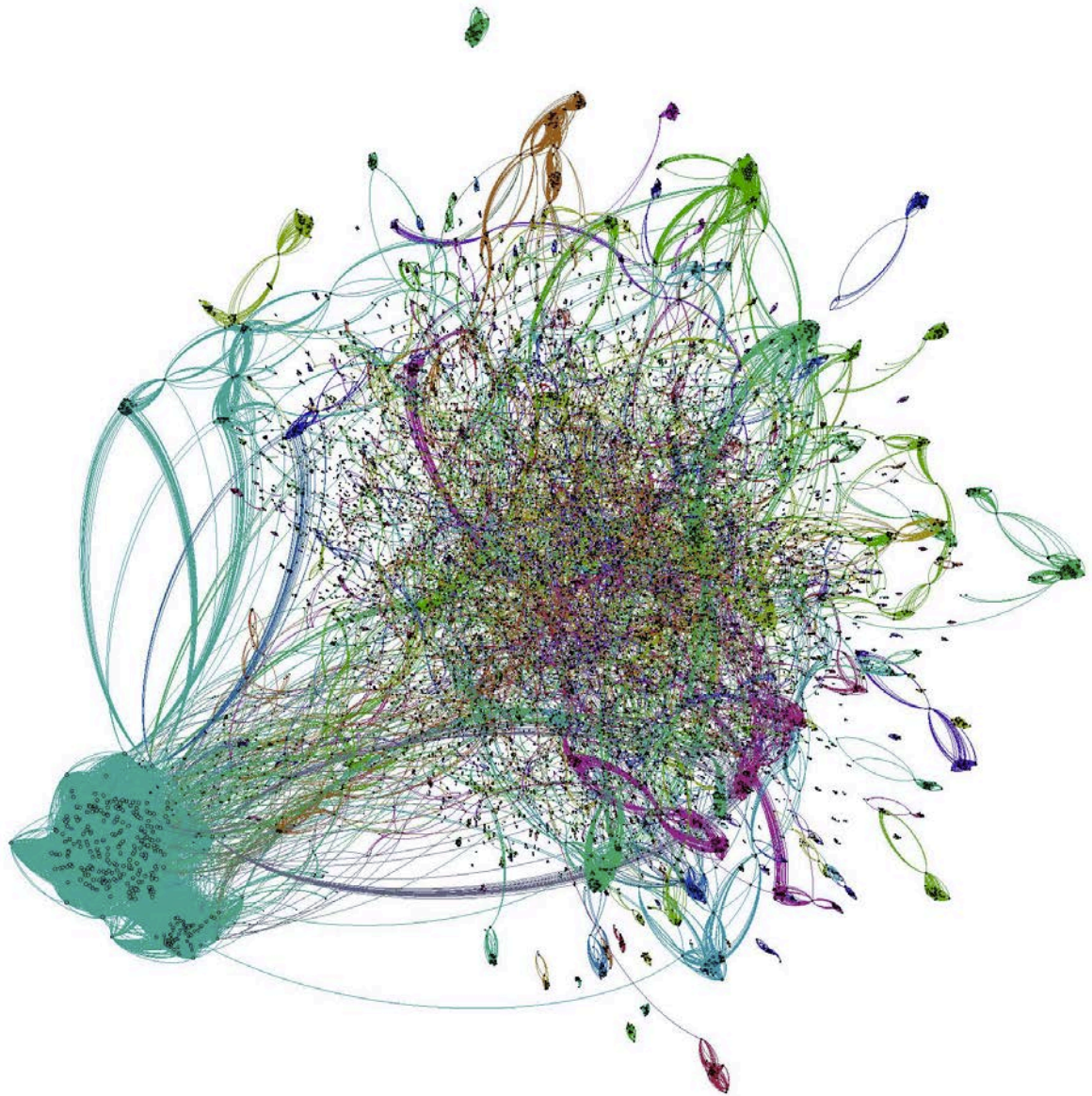

**Figure 1.** Structure of the collaboration network of Physicists on ArXiv in 2002 consisting of 20,464 nodes and 86,237 edges.

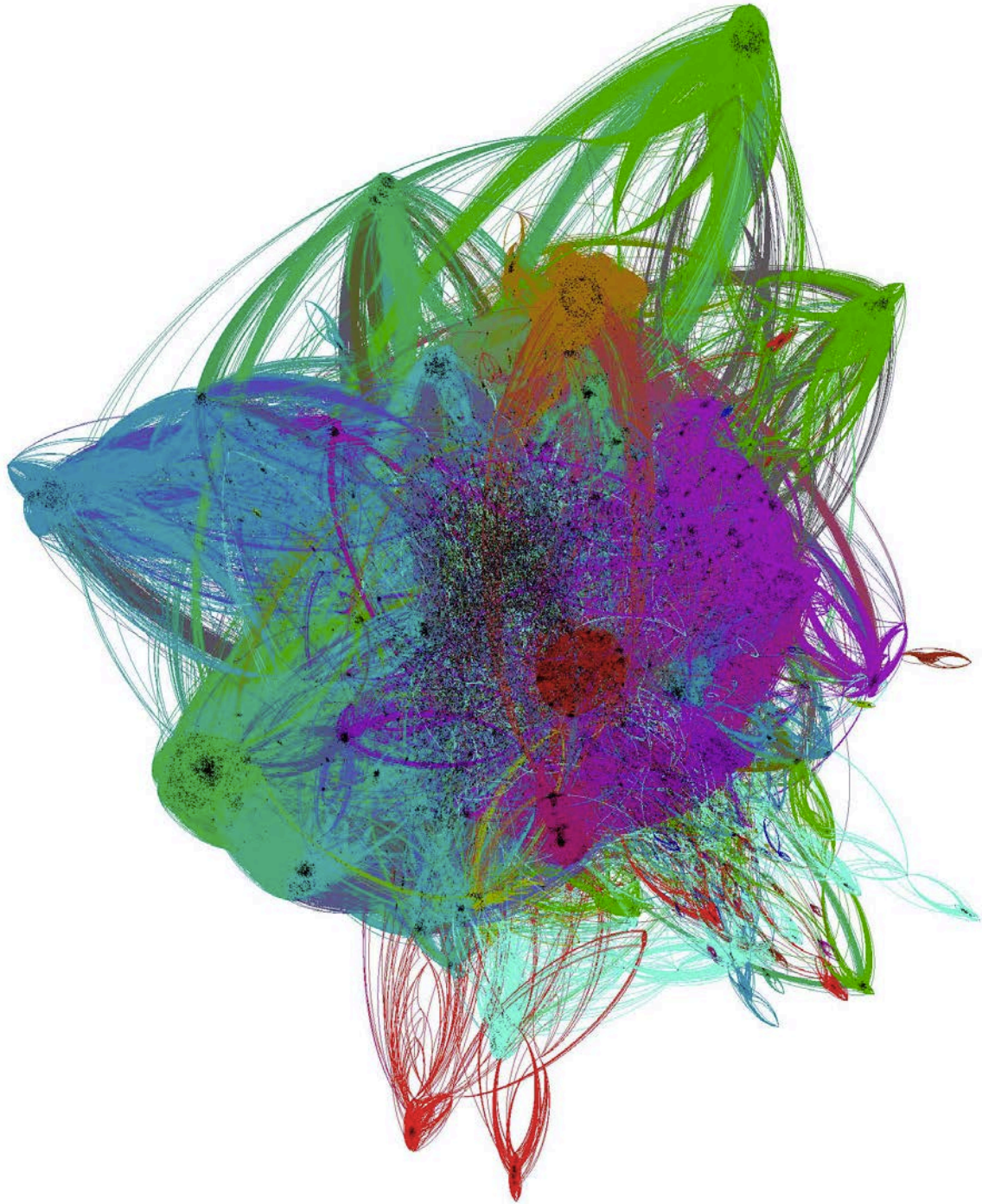

**Figure 2.** Structure of the collaboration network of Physicists on ArXiv in 2011 consisting of 64,785 nodes with over 1 million edges.

### 5.2. Operationalization of Variables

The measurement of scientific success and failure is at the core of our paper. Resting on the assumption that publications are a necessary condition for every scientific career, we first distinguish the articles uploaded to ArXiv between unpublished and published articles. In the latter case, we assume differences in quality and reputation of the journals. It makes a huge difference whether a paper appears in a high impact journal like *"Review of Modern Physics"* or in rather peripheral journals like *"Doklady Physics"*. To assign the journals correctly, we use Thomson Reuters' InCites Journal Citation Reports (ThomsonReuters, 2016). To assess the ranking of journals as undistorted as possible, we sorted the journals by subdisciplines and by five-year citation rate. In this way, we take the (significant) different sizes of communities in Physics into account which are reflected in the number of journals available per subdiscipline. We only consider the leading journals of each specialization.[3] For convenience, we defined high impact journals as the ten journals with the highest outreach in terms of five-year citations of each subdiscipline resulting in the definition of 83 high impact journals out of 456 journals present in our data set. Originally, there have been 90 high impact journals for a total of 9 subdisciplines, but the number is smaller than 90 since seven of them were listed as high impact journal in two subdisciplines. To avoid overlaps we included the respective high impact journals in only one of the two subdisciplines.

The required information was retrieved from the "journal reference" field of ArXiv where all external publications are recorded. The update of the journal references is managed by the authors themselves, but checked and completed in cooperation with large institutional libraries like the Stanford Public Information Retrieval System, one of the largest databases in Physics, but also in cooperation with journal publishers (ArXiv, 2016c; 2016d). Because we do not know the basic population of all non-published articles but need to assess the quality of the information provided, we randomly selected several dozens of papers that have no journal reference and checked the scientific repositories of Google Scholar and Web of Science for any entries. We found none, thus the journal reference field seems to be a resilient indicator. In addition, the relative share of success and failure hovers in a relatively narrow range and is almost equally split (Figure 3). Its fluctuation is no more than 6 percent during our 10-year period of observation, which makes us confident regarding the validity of the dependent variable.

---

[3] An overview of the subdisciplines and journals is provided in appendix A.

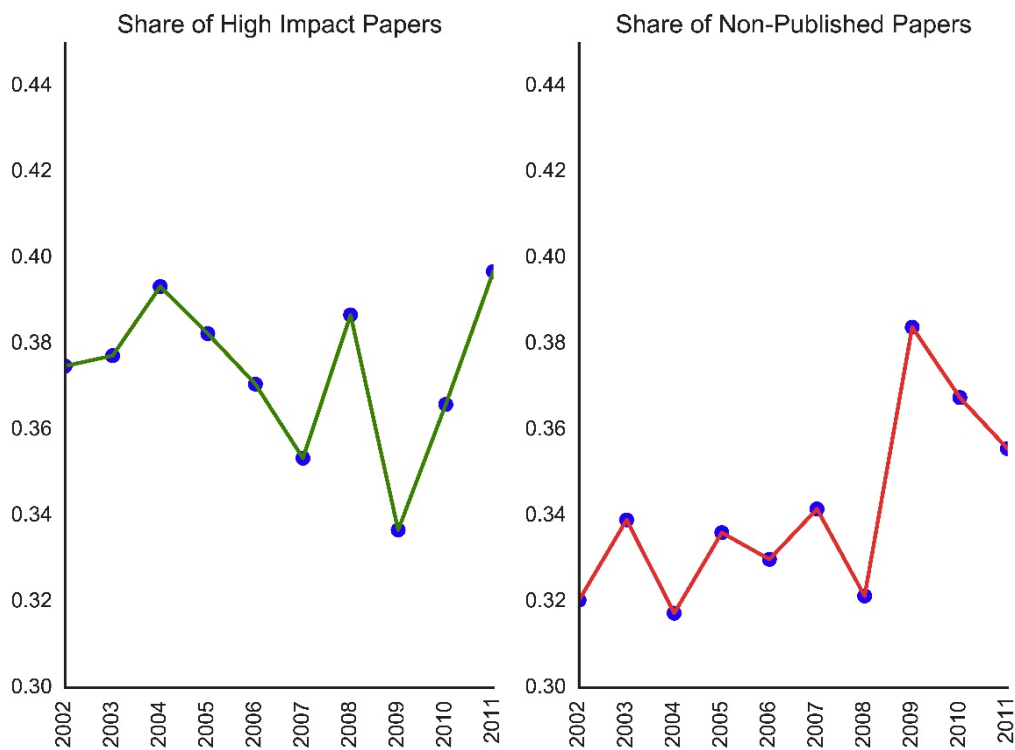

**Figure 3**: Share of high impact papers and of non-published papers.

*Note*: Share of papers published in high-impact journals of all uploaded ArXiv papers in Physics depicted in green. Share of papers not published in any peer-reviewed journals after uploaded on Arxiv is depicted in red.

We now turn to our key question: what individual network properties influence the successful share of papers, i.e. papers being published in high impact journals? Since we are using parametric multilevel event-history analysis, the dependent variable must be binary coded, with a value of one for being published in a high impact journal and zero in all other cases. To test the influence of nodes' position and disciplinary background empirically, we operationalize the hypotheses proposed above as follows:

**H1a**: The number of ties is the degree of each author $d_i$, weighted by the number of authors per paper. The weighted degree can be easily derived for *n* vertices as the row sums from adjacency matrix $\boldsymbol{A}^t$.

**H1b**: Persistent ties are ties that last at least two consecutive years, i.e. the intersection of $\boldsymbol{A}^t$ and $\boldsymbol{A}^{t+1}$. The results are treated as binaries, thus *i* and *j* have or have not collaborated at least once in consecutive years. The number of persistent ties $P_i$ of scientist *i* is therefore simply the number of ties that already existed in the previous year, which enhances the interpretability of the measure.

**H1c**. The social capital of an author is operationalized by his or her eigenvector centrality (Bonacich, 1972; Burris, 2004). Newman (2004) has shown that the correct generalization of eigenvector centrality in a weighted network is still the leading eigenvector of the adjacency matrix with $EV_i = \lambda^{-1} \sum_j A_{ij} x_j$.

**H1d**. The proposed measure to identify structural holes is "network constraint" (Burt, 1995; 2004). It follows the opposite logic of broker positions, i.e. constrained network positions span fewer structural holes assessing whether researchers are concentrated directly or indirectly on single contacts. Innovation, flow of ideas and, in our case, scientific originality should therefore have a negative association with network constraint. It can be calculated by using $c_{ij} = \left(p_{ij} + \sum_q p_{ij} p_{qj}\right)^2$, with *i,j* and *q* being different scientists and $p_{ij}$ being the proportion of all of *i*'s relations (i.e. time and energy) directly invested in the connection with *j* and the indirect investment of both *i* and *j* spent proportionally on their common relationships with *q* ($\sum_q p_{ij} p_{qj}$) representing redundant information for both.

**H2a/b:** Interdisciplinarity is measured on an ordinal scale by combining disciplines in order to bridge intellectual boundaries. We distinguish between papers from different disciplines (e.g. Engineering and Physics, *H2a*) and collaborations between different specializations within Physics (e.g. Computational and Chemical Physics, *H2b*). The reference category is represented by research projects that address only a single disciplinary topic.

### 5.3. Network Positions in Event Studies

In addition to positional measures stemming from social network analysis, we use a two-level event history analysis to model the influence of each position in the co-authorship network on publication success (Steele, 2008). The multi-level approach is necessary because every contribution is nested within an author. The author's centrality for each contribution is found on the first level (i.e. the paper), the author itself on the second level of analysis. This approach results in a parametric, repeated event model as used by Windzio (2006, p.176) and by De Nooy (2011) for network effects. Hereby, we apply a random intercept that follows a certain deviation from an author-specific mean as to the chance of getting a paper published in high impact journals for each paper uploaded at ArXiv.org by a single author.

An event *m* is denoted as any type of publication in a Physics high impact journal with $m = \{1, ..., M\}$. It is assumed that every uploaded paper has an independent chance to witness an event, nested within distinguishable authors. Therefore, the contributions uploaded to ArXiv constitute the *population at risk* for being published (Singer and Willett, 2003, p. 329), while the authors are a source for unit-specific variance in the chance for getting published and are modeled in line with de Nooy (2011, p. 33) as a random effect. The population at risk usually diminishes over time due to event occurrence, dropout or unobservable event times. Therefore, being at risk must be understood conditionally as being part of the population at risk at a time $t_n$, provided that one did not witness an event at $t_{n-1}$. In our case, time is constructed as discrete intervals in years with the intervals [0; t$_1$),[t$_1$;t$_2$),...,[t$_{n-1}$;t$_n$),[t$_n$;∞).

Being published is not the only possible outcome. Many uploaded papers are not published at all (Figure 3). In this case, we call this non-event *right-censored* ($C$) at a fixed time $t_n$ ($C_{t_n}$) (Scott and Kennedy, 2005, pp. 418 – 419). Authors *k* and papers *p* that enter the risk-set at a later time have higher chances to be right-censored – that is to witness no event at all. Following Fine and Grey (1999), it is assumed that right-censoring does not have an effect for a paper being at risk at times $t_r > C_{t_n p}$, with $t_r$ as an observation *r* being at risk after the censoring time of another paper $C_{tp}$ ($p \neq r$). In other words, the paper *p* which entered the population at risk after paper *r*, cannot retroact on the chances of a paper that entered the risk set before being published. We can interpret our findings as chances for being published in a high impact journal in a certain year after uploading it onto ArXiv, provided that the paper was not published the year before. Depicted in odds, the risk (or hazard) to witness a publication in a high impact journal nested within the author on a discrete time scale is then written as:

(1) $h_{kp}(t) = \frac{e^{\alpha(t)+\beta x_{kp}(t)+u_k}}{1+e^{\alpha(t)+\beta x_{kp}(t)+u_k}}.$

Here, $\alpha(t)$ is a constant of change over time. The constant represents the power of an exponential distribution, a log logistic distribution, a Weibull distribution or gamma distribution. The distribution itself is chosen manually as a consequence of the shape of the hazard-function. Values of less than one indicate decreasing chances of getting a paper published over time, while values greater than one indicate increasing chances for a paper to get published over time. $\boldsymbol{\beta x}_{kp}(t)$ represents a vector of covariates, in our case the network measures (weighted degree, eigenvector, constraint, persistent ties and diversity) of the article for each contribution of author *k* for a paper *p*. The random effect for every author *k* is assumed to follow a normal-distribution with a mean of zero and is denoted as $u_k \sim N(0, \sigma_u^2)$.

Furthermore, we need to define a time metric as well as to identify an endpoint of the time of observation in order to calculate the hazard of event occurrence. The crucial feature for the implementation of event-history analysis models is the date of the ArXiv and journal publication. Since the ArXiv data do not include the month of publication, we decided to use years as timeframe. To take into account that papers filed for publication toward the end of the year have smaller chances of being published in that very year, we only consider articles uploaded until the 1st September of each year for establishing values for the respective year.

Due to the data structure and the limitations of the publication data, we had to choose a censoring time. We decided to use the end of 2014 as end of the observation period. To clarify the meaning of this concept of time, imagine two papers that were uploaded in ArXiv in the years 2006 and 2011. In both cases, the year of the upload represents the time t = 0, or initial time. If both were published in the same year, the event time would be t = 1, since an event cannot occur without time elapsed. If the papers were to be published in the following year, time would be t=2 and so forth. Hence, the first could theoretically remain at risk for eight years, while the second one could remain at risk for three years only (t=9 versus t=4). If the first paper had been published before the fourth year (t≤5), it would drop out of the risk set in the fifth year (t=6). In our case, time is denoted as the period from uploading the paper on ArXiv to its publication in a high impact journal.

To compare the goodness of fit of each calculated two-level parametric event-history model, we scrutinize the logarithmic Pseudolikelihood, the Akaike Information Criterion (AIC) and the Bayesian Information criterion (BIC). The Pseudolikelihood function is used to estimate the probability that a vector of variables **X = X₁,X₂, …,Xₙ** corresponds to a vector of values **x**: It is assumed that vectors **X₁,** to **Xₙ** are conditionally dependent, denoted as $\{X_\tau, X_\varphi\} \in E$.[4]

(2) $Loglikelihood = \sum_i \log PR(X_\tau = x_\tau | X_\varphi = x_\varphi \; for \; all \; \{X_\tau, X_\varphi\} \in E)$.

The AIC and BIC are based on the likelihood function and are usually utilized for the comparison between non-nested models with different numbers of observation and different numbers of covariates (Singer and Willett, 2003, pp. 121 – 122). The AIC is calculated as follows:

(3) $AIC = -2[Loglikelihood - X_q]$.

---

[4] The presented loglikelihood function is a generalized form. For a more sophisticated approach and calculation of the loglikelihood function for interval censored data, see Banerjee and Sen (2007).

That is -2 multiplied by the loglikelihood penalized for the number of covariates $X$ used for the respective model $q$. The BIC furthermore includes a scale factor, which is the sample size included for each model. In our case, BIC is calculated as:

(4) $BIC = -2[Loglikelihood - \boldsymbol{M}_{qt0} * \boldsymbol{X}_q]$,

where $M$ is the population at risk at the initial time $t_0$ for each model ($q$) multiplied by the number of covariates to adjust for change in the number of observations.

## 5.4. Covariates

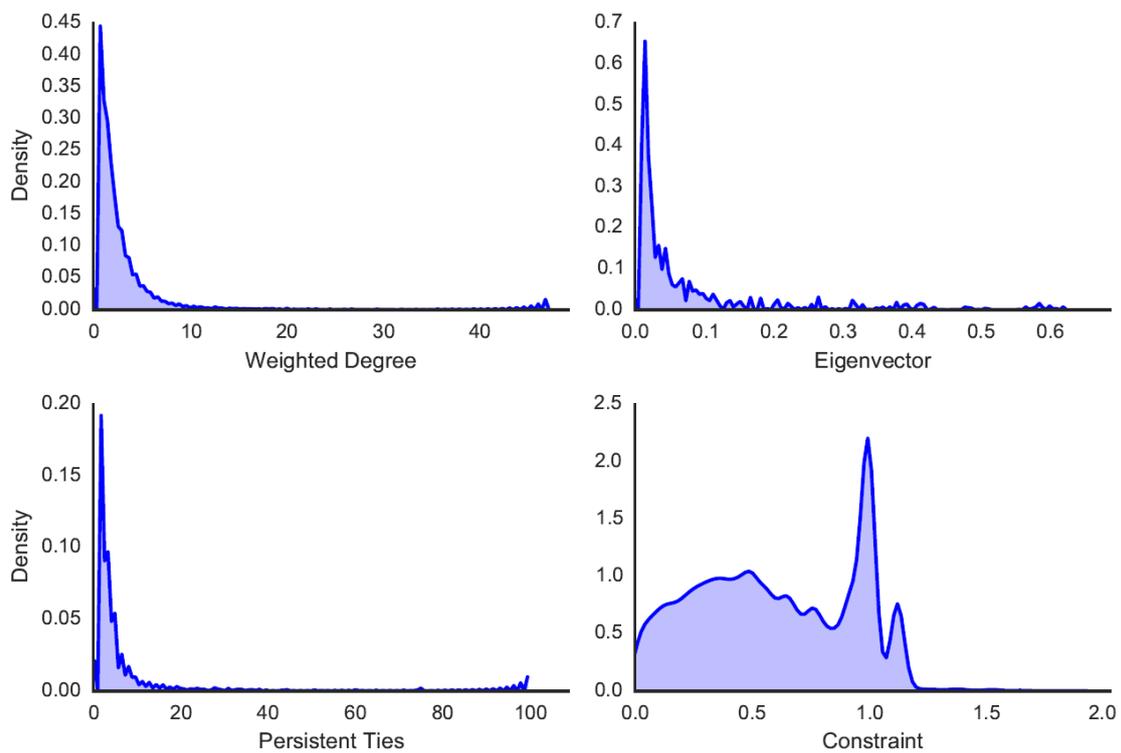

**Figure 4**: Kernel Density Function for weighted degree, eigenvector centrality, number of persistent ties and constraint.

*Note*: The probabilities are calculated per unit (in analogy to other densities like mass per unit volume) and thus the shaded areas of the curves sum up to 1. For instance, the probability of values hovering between $x_1$ and $x_2$ is the area below the curve between those two points.

Normally, covariates in event-history analysis are continuous or dichotomous as in ordinary regression techniques and are interpreted alike. Hence, they are interpreted linearly and have to be increasing in linear terms. Most positional network measures, however, follow power-law distributions (see Figure 4). To facilitate the interpretation (especially of the interaction effects) of the highly right-skewed distribution, we used two cut-off points at the median and the 90%-percentile for each measure and split the data in three categories (Gelman and Park, 2009). We have to use categorized covariates in order to keep the units homomorphous within the interaction effects. To this end, we calculated the respective cut-points for each variable of each year. This results in the categorization listed in table 1.

**Table 1.** Cut-Off Points for the weighted degree, eigenvector centrality, persistent ties and constraint.

| Year | Weighted Degree | | Eigenvector Centrality | | Persistent Ties | | Constraint | |
|---|---|---|---|---|---|---|---|---|
| | Median | 90%-perc. | Median | 90%-perc. | Median | 90%-perc. | Median | 90%-perc. |
| **2002** | 1.50 | 4.53 | $5.17 \times 10^{-9}$ | $8.75 \times 10^{-6}$ | 2 | 6 | 0.74 | 1 |
| **2003** | 1.50 | 4.50 | $3.55 \times 10^{-9}$ | $7.21 \times 10^{-6}$ | 1 | 5 | 0.66 | 1 |
| **2004** | 1.50 | 4.03 | $3.36 \times 10^{-10}$ | $1.86 \times 10^{-6}$ | 1 | 5 | 0.66 | 1 |
| **2005** | 1.50 | 4.98 | $1.05 \times 10^{-8}$ | $1.15 \times 10^{-4}$ | 1 | 6 | 0.63 | 1 |
| **2006** | 1.54 | 4.96 | $3.17 \times 10^{-10}$ | $1.80 \times 10^{-5}$ | 1 | 6 | 0.60 | 1 |
| **2007** | 1.42 | 4.43 | $5.92 \times 10^{-10}$ | $3.53 \times 10^{-6}$ | 1 | 6 | 0.59 | 1 |
| **2008** | 1.50 | 5.00 | $3.53 \times 10^{-9}$ | $2.25 \times 10^{-6}$ | 1 | 7 | 0.58 | 1 |
| **2009** | 1.58 | 5.33 | $3.90 \times 10^{-7}$ | $1.34 \times 10^{-3}$ | 2 | 8 | 0.54 | 1 |
| **2010** | 1.6 | 5.78 | $5.29 \times 10^{-6}$ | $1.36 \times 10^{-3}$ | 2 | 9 | 0.50 | 1 |
| **2011** | 1.74 | 5.92 | $5.89 \times 10^{-6}$ | $0.62 \times 10^{-4}$ | 2 | 10 | 0.41 | 1 |

Furthermore, we implemented interaction effects between constraint and weighted degree as well as with eigenvector centrality. In this way, the level of constraint is examined in respect of a (relative) high, medium or low number of relationships. To study the differential effects of constraint at the center, the semi-periphery and the periphery of a network, we use the interaction effect with the eigenvector centrality. This allows us to check whether central positions in the co-authorship network of Physicists are socially closed or not and whether social closure has distinct effects on different network positions. Finally, we calculated an interaction effect between constraint and diversity. Following *H3c/d* we would expect Physicists with low constraint to be subject to more influences from outside their own (sub-)discipline. According to Burt (2004), less constraint Physicists should have access to more diverse topics. Presumably, the more diverse the topics are, the higher are the chances for original research. In turn, the higher the chances to conduct original research, the higher will be the chances to publish a paper timely in a high-impact journal. The causal links and their relation to the hypotheses are depicted in Figure 5.

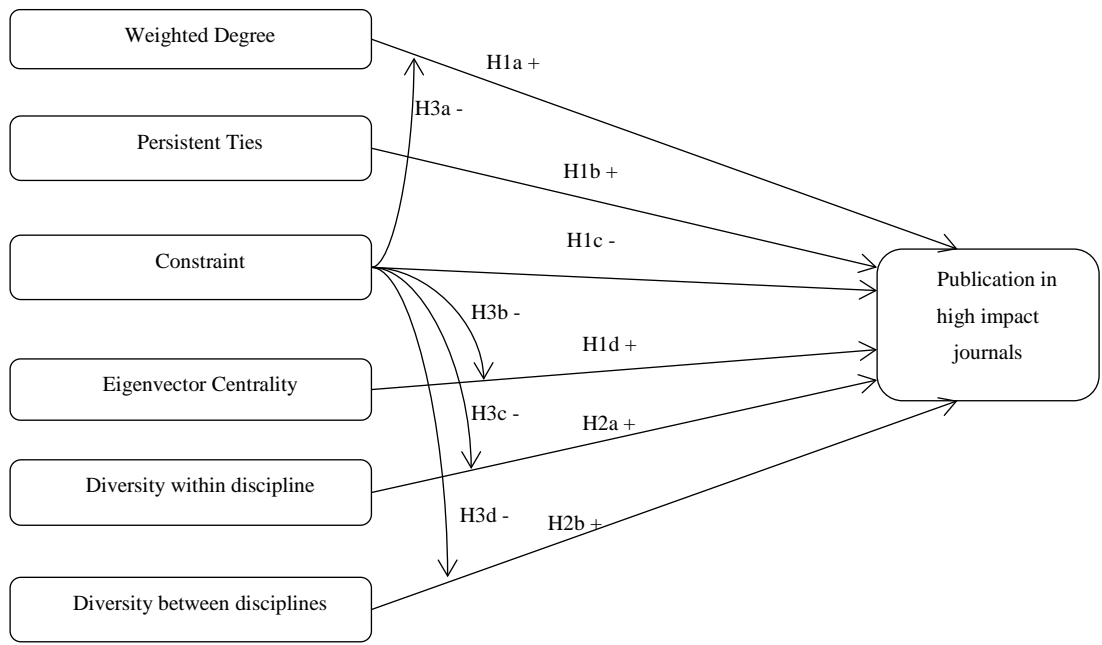

**Figure 5.** Overview of predicted effects of hypotheses.

Before constructing the models of the event-history analysis, we test what type of distribution best fits the probability function of getting a paper published in high-impact journals over time provided that the article was not published before or right censored. To do so, we used a likelihood ratio test, which clearly suggested a Weibull-distribution (Table 2).

**Table 2.** Log Likelihood Test of different parametrical distributions.

| Exponential Distribution | Gamma Distribution | Weibull Distribution | Lognormal Distribution |
|---|---|---|---|
| -321,747.12 | -292,500.04 | -165,235.26 | -297,701.47 |

## 6. Positional Effects on Scientific Success in Collaboration Networks

Our final event-history dataset contains 245,432 uploaded ArXiv papers by a total of 108,348 different authors. The mean number of papers uploaded is 2.3 and the authors accomplish 88,004 publications in high impact journals (35.86%), while 94,752 (38.61%) papers were not published at all. The mean duration of a paper being at risk is 2.14 years with a minimum of less than one year at risk and a maximum of 12 years at risk for the right censored cases that were uploaded in 2002 and censored at the end of 2014 (see also Figure 3).

We utilize four different models. In the first model, the classes for weighted degree, eigenvector-centrality and persistent ties are included. The second model consists of the measures for partial and full diversity. The third model combines the covariates of models 1 and 2, and introduces constraint. Finally, the fourth model adds interaction effects. Note that the reference category for every single variable included in the models is the class with values from 0 to the median values as reported in Table 1.

The value parameter λ (0.8022) indicates a modest decline in the probability to publish as time goes by. The author-specific variance indicates that the chance for getting a paper published in time for an author varies significantly by 0.3820 time-units that is a variation of approximately four months from upload to publication. Because the values of the parameter λ and the author-specific variance do not change on a larger scale in the subsequent models 2, 3 and 4, we will not discuss these in the following sections.

In the first model, we observe no significant differences of having medium or high numbers of ties in comparison to lower numbers of relationships indicated by weighted degrees. Nevertheless, in line with Lin (2002) and Bourdieu (1986) we see that having high volumes of social capital (indicated by eigenvector centrality) is crucial for being published timely in a high impact journal in Physics. The chances of getting a paper published provided that it was not published before is 30.44% higher if the Physicist belongs to the medium category and even by 42.63% if the Physicist belongs to the highest group of eigenvectors.

Yet, it is not only relevant to know other influential Physicists in the network but also to have persistent ties. Having at least one lasting companion increases the chances to publish prominently by 39.21%. However, the effect for being part of the 90%-percentile in numbers of strong ties has a slightly smaller effect. If a Physicist belongs to this category, he or she has a 30.19% higher chance of being published in a high impact journal compared to Physicists without durable ties. This finding is consistent with our assumption that meaningful ties need time to evolve and also much additional time to be maintained. The findings support results provided by Burt (2002), Dahlander and McFarland (2013) and Leahy and Cain (2013). However, it seems that only relatively few persistent ties are sufficient to achieve this effect.

When we control the results for diversity and constraint in model 3 as well as for interaction effects in model 4, we establish a significant effect for Physicists with a large number of collaborations. The findings for Physicists with medium numbers of degrees are mixed. The medium category is significantly decreasing the chances for getting a paper published in model 3, but increases when controlled for interaction effects. Therefore, the findings suggest a *decline of hypothesis 1a in its generality*. As could be observed with persistent ties, it seems to be unnecessary in terms of publication success to possess a very high number of collaborations. Instead, a rather modest number of persistent ties or weighted degrees (i.e. within the medium category) enhances a paper's chances in a high impact journal.

The other positional effects do not diminish when incorporating interaction effects in model 4, thus *hypotheses 1b and 1c cannot be declined*. Having at least a modest amount of persistent ties and possessing many collaboration partners with high volumes of social capital significantly increases the likelihood of placing publications in prominent journals. In fact, the impact of having higher levels of eigenvector centrality in comparison to the lowest category is rising by 116.23%, while the effect of being part of the 90%-percentile of eigenvector-centrality rises even from 42.63% to 331.76% when controlled for interactions.

The second model addresses the influence of different grades of interdisciplinarity of papers on their publication in high impact journals over time. It shows that collaborating with colleagues from other subdisciplines of Physics *lowers* the chances of getting a paper published in a high impact journal by 37.19%. Uploading a working paper with colleagues from a discipline outside Physics also *decreases* the chances for publication in a high impact journal by 16.43%. This finding contradicts our hypothesis that working in interdisciplinary teams leads to a higher chance of being published and, in this sense, of being successful (see Figure 6). Similar to most effects of model 1, the influence of partial and full diversity remains stable in models 3 and 4. *Hypotheses 2a and 2b are therefore rejected*. We cannot find a net effect of interdisciplinary collaborations on scientific success in Physics.

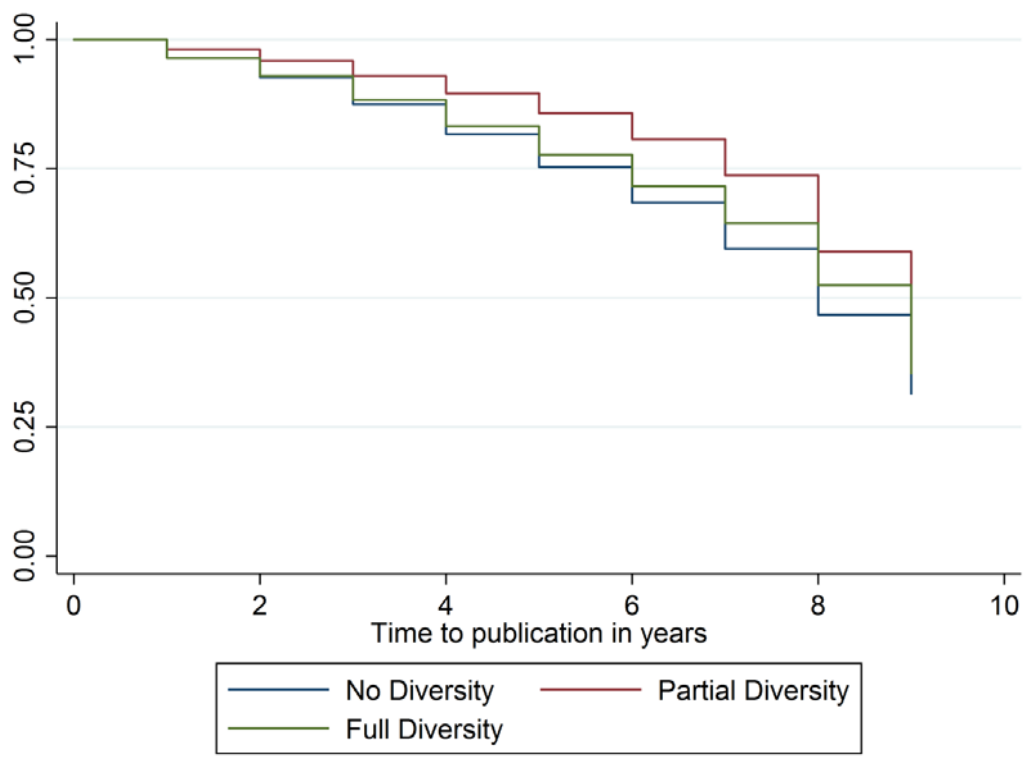

**Figure 6:** Kaplan-Meier estimations of survival probability of a paper categorized by diversity over time.

*Note*: The curves represent the "survival probability", i.e. the time elapsing until a working paper uploaded on ArXiv gets published in a high impact journal. Therefore, the lower the curves the more successful are authors in the respective category.

**Table 3.** Estimates of the two-level parametric event-history analysis.

|  | Model 1 | Model 2 | Model 3 | Model 4 |
|---|---|---|---|---|
| Weighted Degree, medium [a] | 0.9986 |  | 0.9793* | 1.3516** |
|  | (0.0095) |  | (0.0097) | (0.1172) |
| Weighted Degree, high [a] | 1.0191 |  | 1.0036 | 0.5501 |
|  | (0.0286) |  | (0.0282) | (0.5789) |
| Eigenvector-Centrality, medium [b] | 1.3044*** |  | 1.2835*** | 2.1623*** |
|  | (0.0110) |  | (0.0112) | (0.1933) |
| Eigenvector-Centrality, high [b] | 1.4263*** |  | 1.4157*** | 4.3176*** |
|  | (0.0269) |  | (0.0269) | (1.3250) |
| Persistent Ties, medium [c] | 1.3921*** |  | 1.3758*** | 1.3750*** |
|  | (0.0158) |  | (0.0156) | (0.0156) |
| Persistent Ties, high [c] | 1.3019*** |  | 1.2386*** | 1.2455*** |
|  | (0.0303) |  | (0.0289) | (0.0291) |
| Partial Diversity [d] |  | 0.6281*** | 0.6368*** | 0.5480*** |
|  |  | (0.0062) | (0.0062) | (0.0237) |
| Full Diversity [d] |  | 0.8357*** | 0.8446*** | 0.8889** |
|  |  | (0.0084) | (0.0084) | (0.0368) |
| Constraint, medium [e] |  |  | 0.9523*** | 1.2692 |
|  |  |  | (0.0087) | (0.1883) |
| Constraint, high [e] |  |  | 0.9302*** | 1.5678 |
|  |  |  | (0.0173) | (1.7189) |
| Constant | 0.0047*** | 0.0059*** | 0.0054*** | 0.0032*** |
|  | (0.0001) | (0.0001) | (0.0001) | (0.0035) |
| $\lambda$ | 0.8022*** | 0.8085*** | 0.8086*** | 0.8088*** |
|  | (0.0029) | (0.0029) | (0.0029) | (0.0029) |
| Level 2-Variance | 0.3820*** | 0.4252*** | 0.3749*** | 0.3714*** |
|  | (0.0085) | (0.0087) | (0.0083) | (0.0083) |
| Log-Pseudolikelihood | -160,951.24 | -160,983.57 | -159,804.03 | -159,655.28 |
| AIC | 321,920.67 | 321,977.14 | 319,634.05 | 319,360.56 |
| BIC | 322,014.17 | 322,029.19 | 319,769.39 | 319,620.83 |
| No. Observations | 245,432 | 245,432 | 245,432 | 245,432 |
| No. Authors | 108,348 | 108,348 | 108,348 | 108,348 |
| No. Events | 88,004 | 88,004 | 88,004 | 88,004 |
| No. Censored | 94,752 | 94,752 | 94,752 | 94,752 |

**Table 3.** Continued.

| | | | | |
|---|---|---|---|---|
| Low constraint × Weighted Degree, low [f] | | | | 0.5449 |
| | | | | (0.5736) |
| Low constraint × Weighted Degree, medium [f] | | | | 0.39866 |
| | | | | (0.4188) |
| Medium constraint × Weighted Degree, low [g] | | | | 0.4397 |
| | | | | (0.4668) |
| Medium constraint × Weighted Degree, medium [g] | | | | 0.3118 |
| | | | | (0.3322) |
| Low constraint × Eigenvector-Centrality, low [h] | | | | 3.1771*** |
| | | | | (0.9771) |
| Low constraint × Eigenvector-Centrality, medium [h] | | | | 1.7887 |
| | | | | (0.8868) |
| Medium constraint × Eigenvector-Centrality, low [i] | | | | 2.8621** |
| | | | | (0.8868) |
| Medium constraint × Eigenvector—Centrality, medium [i] | | | | 1.8632 |
| | | | | (0.5993) |
| Low constraint × Partial Diversity [k] | | | | 1.2671*** |
| | | | | (0.0573) |
| Medium constraint × Partial Diversity [k] | | | | 1.0651 |
| | | | | (0.0488) |
| Low constraint × Full Diversity [l] | | | | 0.8975* |
| | | | | (0.0391) |
| Medium constraint × Full Diversity [l] | | | | 1.0105 |
| | | | | (0.4427) |
| Constant | 0.0047*** | 0.0059*** | 0.0054*** | 0.0032*** |
| | (0.0001) | (0.0001) | (0.0001) | (0.0035) |
| λ | 0.8022*** | 0.8085*** | 0.8086*** | 0.8088*** |
| | (0.0029) | (0.0029) | (0.0029) | (0.0029) |
| Level 2-Variance | 0.3820*** | 0.4252 | 0.3749*** | 0.3714*** |
| | (0.0085) | (0.0087) | (0.0083) | (0.0083) |
| Log- Pseudolikelihood | -160,951.24 | -160,983.57 | -159,804.03 | -159,655.28 |
| AIC | 321,920.67 | 321,977.14 | 319,634.05 | 319,360.56 |
| BIC | 322,014.17 | 322,029.19 | 319,769.39 | 319,620.83 |
| No. Observations | 245,432 | 245,432 | 245,432 | 245,432 |
| No. Authors | 108,348 | 108,348 | 108,348 | 108,348 |
| No. Events | 88,004 | 88,004 | 88,004 | 88,004 |
| No. Censored | 94,752 | 94,752 | 94,752 | 94,752 |

**Reference categories:** [a] low weighted-degree centrality; [b] low eigenvector-centrality; [c] low number of persistent ties; [d] no diversity; [e] low constraint; [f] low constraint and high weighted-degree centrality; [g] medium constraint and high weighted-degree centrality; [h] low constraint and high eigenvector-centrality; [i] = medium constraint and high eigenvector-centrality; [j] high constraint and partial diversity; [k] high constraint and full diversity

*Note*: Standard errors in parentheses

In *model 3,* the values for *constraint* are included. Hypothesis *1d* predicted that a higher constraint in scientific relationships decreases the chances to publish a paper prominently. The reported results support that assumption. Being more constrained than the average lowers the chances for a Physicist to succeed in being published by 4.77%, while being highly constrained diminishes the chances by 6.98%. However, if we take the *interaction coefficients* in *model 4* into account, effects of constraint change. It becomes insignificant if central positions in the collaboration network are considered while the value of eigenvector centrality increases immensely. The combination of both effects underscores that being central *and* having large amounts of social capital and persistent ties in a network increases the likelihood of scientific success tremendously. In the case of (semi-)peripheral positions, indicated by the medium and low category of the eigenvector, being constrained works quite differently. Physicists with both low social capital and low constraint values have 217.71% higher chances of getting a paper published than Physicists with high constraints in the same social capital category. The same effect is at work for the medium constraint category. Here, Physicists with low constraint have 186.21% higher chances to publish their findings in a high impact journal, provided that these were not published the year before as emphasized in Figure 7. However, the effects of constraint are not significant for medium categories of social capital. It is only when a scholar is not or very little constrained that chances rise to get a paper published in a high impact journal in contrast to highly constrained and (semi-)peripherally located scholars. These findings suggest that positions with low resources in terms of prestige and reputation can be very helpful collaboration partners if they are not highly constrained. Turning the argument around, scientists with few prestigious partners should mind diverse collaboration partners as non-redundant information sources. Therefore, the findings suggest that *hypothesis **1d** is rejected*, while *hypothesis **3b** is partially supported*.

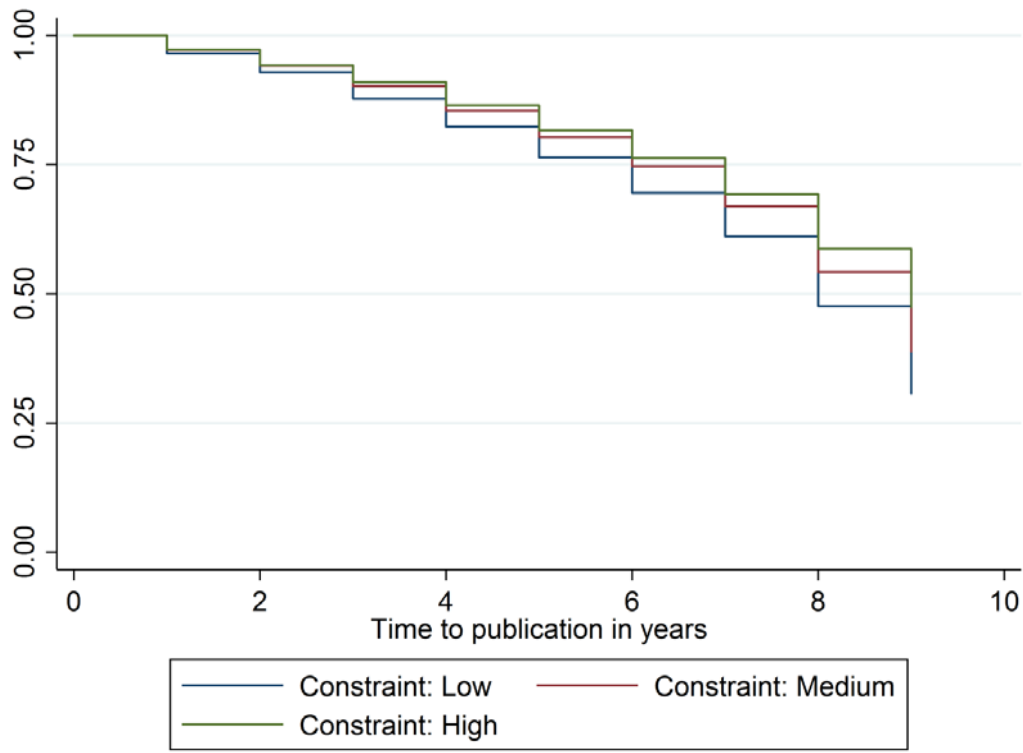

**Figure 7:** Kaplan-Meier estimations of survival probability of a paper categorized by constraint over time.

Finally, we turn to the effect whether brokers spanning structural holes connect "good ideas" (Burt, 2004), i.e. whether partners stemming from multidisciplinary backgrounds are enhancing the probability of publishing in Physics high impact journals. The results in model 4 show that having lower levels of constraint increases the chances of getting a paper published by 26.71% as compared to subdisciplinary collaborations with high levels of constraint (Figure 8). Only researchers in brokerage positions can therefore utilize collaborations with colleagues from different subfields, which partly confirms hypothesis **3c**. Nevertheless, in the case of full disciplinary collaboration even that does not apply. If Physicists collaborate with colleagues from other disciplines, e.g. Mathematics or Engineering, their chances of getting a paper published in Physics high impact journals diminish by an additional 10.25% at a level of 5%-significance. These findings indicate that being able to transcend (sub)disciplinary boundaries does not necessarily make it more likely to get a paper published in high impact journals and being recognized by colleagues (Figure 9). On the contrary, being highly specialized whilst still being at the center of the network and having a sufficient amount of trusted colleagues is the key to success. Therefore, *hypothesis 3d must be declined*.

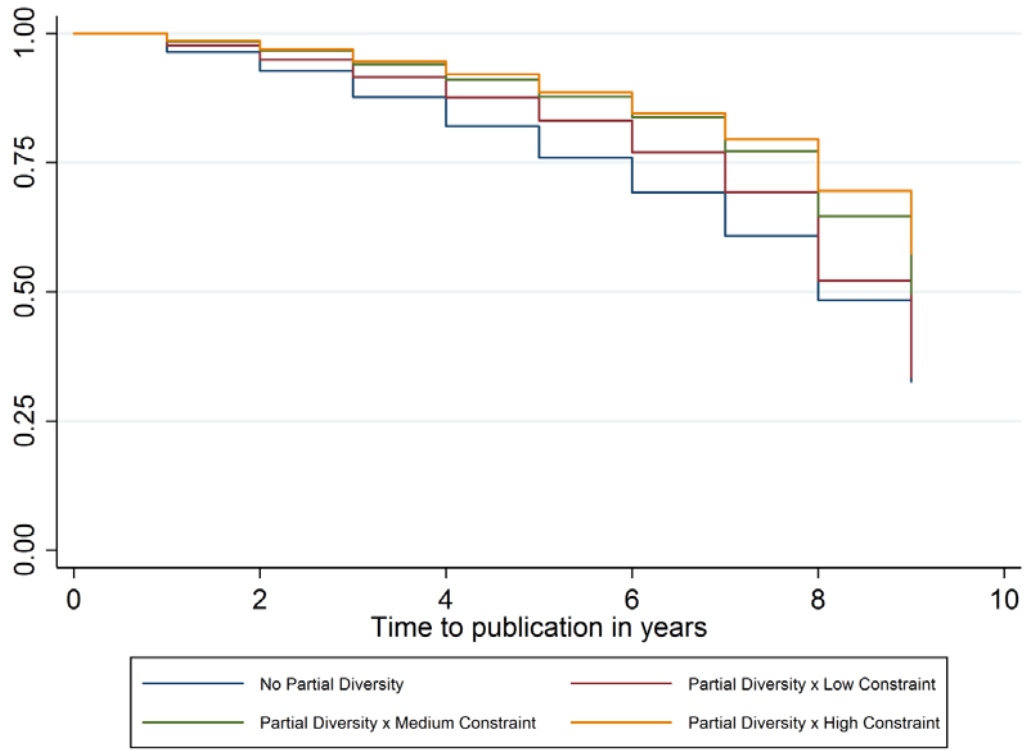

**Figure 8:** Kaplan-Meier estimations of survival probability of a paper categorized by partial diversity * constraint over time.

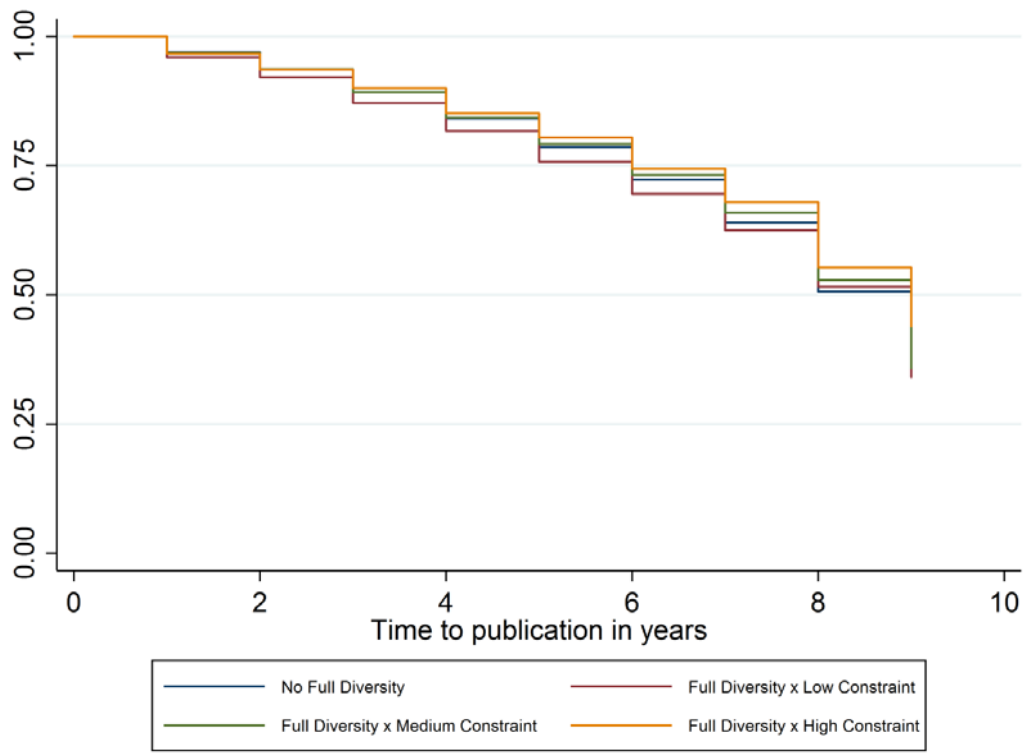

**Figure 9:** Kaplan-Meier estimations of survival probability of a paper categorized by full diversity * constraint over time.

## 7. Discussion

The results presented in this paper reveal new insights into network effects that influence scientific success in Physics. The strong necessity to collaborate in this research community makes it an important case to investigate what network positions increase or decrease the probability of papers being successful and placing them in high impact journals. The ArXiv data allows us not only to look at the successful papers and their authors, but also to consider the rarely observed other side of the coin, i.e. failure in terms of papers that are not published at all or in peripheral journals only. We discover several factors associated with the "input" dimension of science as criticized by Münch (2014). Three findings stand out in particular: (1) Constrained relationships hamper success, though not for all network positions, (2) persistent ties are relevant for publication success, and (3) interdisciplinarity does not necessarily increase success in terms of getting a paper published.

First, constrained scientific relationships show the expected negative influence on success. This is in accordance with Abbasi et al. (2012) who also identify a positive effect of brokerage positions on scholars' research performance in Information Science. In contrast, Bordons et al. (2015) report no positive effect of bridging roles on research output. However, their sample concentrates exclusively on Spanish scientists and therefore reflects this country's structure only. In contrast, our data emphasizes that positions with low constraint (i.e. brokers) indeed increase the publication probability in Physics, though only under certain conditions. Including interaction effects, the negative role of constraint relations is no longer significant. Yet, there is a very strong and highly significant influence on scientific success if bridging positions are endowed with low or medium levels of social capital. Low constraint, in other words, works best when it is *not* combined with high eigenvectors. One reason is that leaders of teams as supposedly well-performing scientists and drivers of collaborations should write most of their papers together with their assistants and students so that their eigenvector will not reach high values (Rawlings et al., 2015). Another part of the explanation concerns a network's periphery in terms of social capital. We suggest that those positions are mostly held by young and, generally, innovative scientists. In fact, if their collaborations are connecting non-redundant information (i.e. low constraint), this enhances the chances of success significantly. In the case of more peripheral and less constrained Physicists, we can therefore argue in line with Uzzi et al. (2013, p.471) that "highly conventional combinations of prior work, with an intrusion of combinations unlikely to have been joined together before", are successful.

From a stratification-oriented perspective, it is possible that these positions are occupied by researchers from non-elite departments. Physicists located at the (semi-)periphery could boost their career by tailoring their research and choosing literally the "right" collaboration partners located at the center of the respective networks. More generally, the interaction effects between eigenvector and constraint in Physics indicate that those measures are not only theoretically complementary (Walker et al., 1997), but should also be statistically considered as interaction effects. Hence, scholars focusing on power asymmetries, social stratification or life-course analysis could benefit substantially if they considered network positions *and* their interactions.

Our second main finding is the effect of persistent ties. It reveals that it is not only important to form ties, but also to maintain them. Emphasizing the importance of stable relationships is consistent with the findings of the case studies conducted by Leahy and Cain (2013), as well as Dahlander and McFarland (2013), Godechot (2016) and Ylijoki (2014, p. 68), who emphasized that "[e]veryone wants to get good partners and avoid bad ones, who complement one's area of competence, bring in new funding sources, open up interesting interdisciplinary research paths, offer synergy advantages, and introduce nice colleagues". The time consuming effort to maintain ties is represented by the threshold effect for medium numbers of persistent ties.

On a theoretical level, these findings reveal that the measurement of weighted degree and eigenvector centrality at a single point in time does not suffice to be interpreted as social capital. Instead, both time and effort to maintain and strengthen different types of ties should be taken into account. This demand is in line with the interpretation of social capital considered by Burt (1995, p. 2004). It may be equally time consuming and difficult to maintain ties directed outside one's own research community, linking, for instance, different epistemic cultures.

The last and probably most important result leads us to the effect of interdisciplinary collaborations. Interdisciplinarity appears often as *panacea* in science (Sa, 2008; Boardman, 2009; Graffkin and Perry, 2009; Viale and Etzkowitz, 2010; Lynn, 2014). However, our findings support a more critical view on this topic in line with Boden et al. (2011). In all models, disciplinary diverse collaborations have a strong negative influence on success. This is connected to our operationalization of successful papers, since the criteria are not publications in, by definition, multidisciplinary journals (especially Nature and Science) but in leading journals of specific research communities. In our view, these journals are decisive for most peers and publications therein form the basis of every scientific career. Uzzi et al.

(2013, p. 471) stated that many "hit papers" follow "highly conventional combinations of prior work". Considering the basic requirement of receiving many citations – i.e. getting a paper published in high impact journals – we see that in Physics particularly collaborations between subdisciplines are deteriorating these success probabilities (Figure 7).

Besides the specific requirements of disciplinary journals, there is another possible explanation for the lower chances of (sub)interdisciplinary projects to be successful. This explanation roots in difficult research conditions, which interfere with the chance to establish and maintain ties. As Jacobs and Frickel (2009) pointed out, the department structure of universities which in itself structures the academic labor market, discourages researchers to collaborate with colleagues of other disciplines. Furthermore, interdisciplinarity and its benefits for society are difficult to recognize and to valuate by funding agencies and are funded less as Hicks (2009) demonstrated in the case of the British Research Assessment Exercise. Similar findings are provided by Bromham, Dimmage and Hua (2016) in the case of Australia. These conditions may hinder collaboration effectively, leading to short term cooperation at best and hindering scholars to establish the persistent ties needed to be successful.

This finding is in accordance with the general notion that research throughout disciplines gets more specialized (Münch, 2014, pp. 222 – 233; Rushforth and de Rijcke, 2015). As has been discussed in regard to persistent ties, the possibilities to cooperate are relatively rare in these cases and tend to be unevenly distributed into the direction of already established research discourses. This effect may cause a linkage between established/center and not-established/outsider, as it has been discussed by Elias and Scottson (1994). The latter ones are typically more constrained and have lesser chances to get published in high impact journals, as Figure 8 underlines for partial diversity and Figure 9 for full diversity. Thus interdisciplinary research in Physics is, by trend, published in medium or low impact journals.

Our findings meet certain limitations in omitted yet potentially influential antecedents. On the one hand, institutional affiliations are not reported resiliently in the ArXiv data. Only 7 percent of all authors specify their home institution, which would have meant a dramatic loss in sample size. The number of cases together with its longitudinal composition, on the other hand, prevented us from merging this information manually. We hope to find a way in the future to include an institutional level in order to ask even more complex "ecological" (McFarland et al., 2014) questions that allow us to explore the micro-macro link between universities and scholars in regard of scientific success. One potential bias, for instance, could be the fact that central positions in the outlined collaboration network are mainly occupied by scholars located in the United States. In this sense, Kragh (2002) and Marginson (2006) suggest that the United States is the well connected, powerful center of science which is able to set trends in research. Therefore, hidden "ecological" effects might be attributed to a global, yet nationally imprinted research culture represented in both the structure of the presented collaboration network and individual chances for publication.

A second empirical limitation of our paper is the omission of demographic attributes. We have no possibility to control, for instance, gender, race, etc. of the Physicists publishing on ArXiv. One potential solution could be to draw a smaller randomized sample that could be investigated for those individual covariates. From a data point of view this problem would be easier to handle than the first limitation, since those attributes do not change over time (in contrast to institutional affiliations).

Another limitation is connected with the argument of Physics as an epistemic culture. The ArXiv platform is by design mostly relevant for Physicists. We are therefore unable to control whether our results are meaningful for other disciplines or whether we discovered network effects specific to the community of Physicists. This could be a worthwhile direction for future research, since disciplines do have other collaboration and publication opportunities and restraints (e.g. Jansen et al., 2010; Whitley, 2000).

Despite these limitations, our contribution highlights the possibility of failure in science, emphasizes the role of social closure depending on central/peripheral positions in the network, debunks the role of interdisciplinarity for scientific success and underscores the importance of trustworthy, persistent ties for the concept of social capital as introduced by scholars like Bourdieu (1986) or Coleman (1990). We also emphasize the connection between time and success, which is embodied in the time elapsing from uploading a working paper to its publication. Our findings regarding the importance of persistent ties that last at least two years present a strong argument in favor of collaboration and against unlimited competition as well as short-term working contracts in academia, as described by Rhoades (2012). Research might be a competition but academia is a collaborative social endeavor to produce new knowledge. An environment that fosters mistrust and leads to an ongoing "battle royal" will cause networks to fracture, leading to the formation of scientific cults while hampering knowledge flows on the structural level, as has been criticized by Abbott (2006) and Turner (2016) in the case of sociology.

Future studies could focus on the relation between network positions, the durability of ties and success in different societal fields. Success and time until publication could also constitute valuable node types in the hypergraph approach taken by Shi et al. (2015) to understand the "fabric of science" in greater detail. Including network positions as conditions for success in other social fields, for instance, the job market, PhD completion or professional sports could also turn out to be fruitful future research endeavors.

## 9. Appendix

The ArXiv-subdisciplines used for assignment of journals to the journal impact factor.

| Biology | Business | Chemistry | Computerscience | Finance | Mathematics | Physics |
| --- | --- | --- | --- | --- | --- | --- |
| Biochemical Research Methods | Business | Analytical Chemistry | AI | Economics | Pure Mathematics | Astrophysics |
| Biochemistry | Business and Finance | Applied Chemistry | Cybernetics | Management | Applied Mathematics | Atomic Physics |
| Biodiversity | | Inorganic Chemistry | Hardware Architecture | | Interdisciplinary Mathematics | Condensed Matter Physics |
| Biology | | Medical Chemistry | Information Systems | | | Fluids and Plasmas |
| Biophysics | | Organic Chemistry | Interdisciplinary Applications | | | Applied Physics Mathematical Physics |
| Biotechnology | | Physical Chemistry | Software | | | Multidisciplinary Physics |
| Evolutionary Biology | | | Theory and Methods | | | Nuclear Physics |
| Mathematical and Computational Biology | | | | | | Particles and Fields and Statistics |